\documentstyle[twocolumn,prl,aps,epsfig]{revtex}
\begin{document}
\flushbottom
\draft
\title{Optical control and entanglement of atomic Schr\"odinger fields}
\author{M. G. Moore and P. Meystre}
\address{Optical Sciences Center and Department of Physics\\
University of Arizona, Tucson, Arizona 85721\\
(July 23, 1998)
\\ \medskip}\author{\small\parbox{14.2cm}{\small \hspace*{3mm}
We develop a fully quantized model
of a Bose-Einstein condensate driven by a far off-resonant pump laser
which interacts with a single mode of an optical ring cavity.
In the linear regime, the cavity mode exhibits spontaneous exponential gain
correlated with the appearance of two atomic
field side-modes. These side-modes and the cavity field
are generated in a highly entangled state, characterized by thermal
intensity fluctuations in the individual modes, but with two-mode correlation
functions which violate certain classical inequalities.
By injecting an initial coherent field into the optical cavity one
can significantly decrease the intensity fluctuations at the expense of
reducing the correlations, thus allowing for optical control over
the quantum statistical properties of matter waves.
\\[3pt]PACS numbers: 03.75-b,03.75.fi,42.50.Vk,42.55-f  }}
\address{}\maketitle
\maketitle
\narrowtext
The recent demonstration of Bose-Einstein condensation in low-density alkali
vapors \cite{BEC1,BEC2} opens up a new paradigm in AMO physics. It is now possible to
generate macroscopic atomic fields whose quantum
statistical properties can in principle be manipulated and controlled, very
much like those of quantum optical fields. One important consideration is 
to determine to which extent the quantum state of a many-particle atomic 
field can be {\it optically} manipulated.
In the single-particle case, the answer to this problem is known to a large
extent. This is the domain of atom optics \cite{Mlynek}, where a number of
optical elements for matter waves have now been developed, including gratings,
mirrors, interferometers, resonators, etc. But these optical elements
manipulate just the atomic field ``density'', or at most first-order coherence
properties. However, Schr\"odinger fields possess a wealth of further
properties past their first-order coherence, including atom
statistics, density correlation functions, etc. In analogy to the optical
case, one can therefore think of ``quantum atom optics'' as
that extension of atom optics where the quantum state of a many-particle
matter-wave field is being controlled, characterized and used in novel
applications.

This Letter presents an analysis of a system where an
optical field is used to manipulate the quantum state of
a matter-wave field. Using a geometry identical to that used in the
collective atomic recoil laser \cite{BonSal94,HemBigKat96,MooMey98b}, and
very similar to
those used for atom interferometry \cite{Berman}, and recoil-induced
resonances \cite{GuoBerDub92,CouGryLou94}, we consider a Bose-Einstein 
condensate driven by a far off-resonant pump laser and coupled
to a single mode of an optical ring-cavity. This results in gain in the
cavity mode, as well as the generation of momentum side-modes
of the BEC, which are assumed to be unbound by any magnetic or optical
trap. The quantum statistical properties of the side-modes
can be strongly manipulated by varying the initial state of the optical cavity
mode. In addition, a strong quantum
mechanical entanglement can develop between the optical and matter-wave
fields, as well as between matter-wave side-modes. The experimental realization
of this system is currently feasible, in view of recent experiments
on the diffraction of condensates by the NIST group \cite{NIST}.

We consider an ultracold sample of bosonic atoms driven by a strong classical
``pump'' and a counterpropagating weak quantized  ``probe'' optical field,
both being far off-resonant from any electronic transition. Under these
conditions the internal atomic degrees of freedom can be adiabatically
eliminated and the matter-wave field is effectively scalar.
The combined Hamiltonian for the atomic and probe fields is
\begin{eqnarray}
\hat{H}&=&\frac{\hbar^2}{2m}\sum_{\bf q} q^2\hat{c}^\dag({\bf q})
\hat{c}({\bf q})
+\hbar ck\hat{A}^\dag\hat{A}\nonumber\\
&+&i\frac{\hbar}{2\Delta}\sum_{\bf q}\left[g\Omega_0 e^{-i\omega_0 t}
\hat{A}^\dag\hat{c}^\dag({\bf q}-{\bf K})\hat{c}({\bf q})
- H.c.\right]\nonumber\\
&+&\frac{\hbar}{\Delta}\left(\frac{|\Omega_0|^2}{4}
+|g|^2\hat{A}^\dag\hat{A}\right)
\sum_{\bf q}{\hat c}^\dag({\bf q})\hat{c}({\bf q}).
\label{H}
\end{eqnarray}
Here, $\Omega_0$ is the Rabi frequency of the pump laser of frequency
$\omega_0$ and momentum ${\bf k_0}$, ${\hat A}$ is
the annihilation operator of the probe field of frequency
$\omega$ and momentum ${\bf k}$, satisfying $[A, A^\dag] =1$, and
${\hat c}({\bf q})$ is the annihilation operator for a ground state atom of
momentum ${\bf q}$, satisfying $[{\hat c}({\bf q}), {\hat c}^\dag({\bf q}')] =
\delta_{{\bf q},{\bf q}'}$. In addition, $\Delta$
is the detuning between the pump frequency and the upper
electronic level closest to resonance,
and $g=d[ck/(2\epsilon_0LS)]^{1/2}$ is the atom-probe coupling constant.
Here $d$ is the atomic dipole moment, $L$ the length of the
ring resonator
sustaining the probe mode, and $S$ the cross-section of that mode in
the region of the atomic sample. Finally, ${\bf K} \equiv {\bf k_0} - {\bf k}$
is the atomic recoil momentum resulting from the absorption of a pump photon
followed by the emission of a probe photon.

The first two terms in Eq. (\ref{H}) are the free Hamiltonians of the atomic
and probe fields, respectively. The remaining terms
correspond to the various processes by which an atom undergoes a virtual
transition under the influence of the optical fields. The first such term
involves the exchange of a photon between the pump and probe
fields, e.g. stimulated absorption of a pump photon followed by stimulated
emission of a probe photon, or vice versa. In coordinate representation,
this term would take the form of the familiar periodic optical potential
generated by the counterpropagating pump and probe lasers fields.
The last two terms in Eq. (\ref{H}) correspond to processes
where a photon is first absorbed and then reemitted into the same field.
These transitions are recoilless, but contribute a cross-phase modulation
between the atomic and optical fields.

Assuming that the initial momentum width of the condensate is
small compared to the recoil momentum $K$, it is reasonable to treat it
as a {\em single mode} atomic field of momentum ${\bf q}=0$.
We furthermore  restrict our discussion to the case $T \ll T_c$,
where $T_c$ is the critical temperature, and assume a large condensate for
which the bare mode $q=0$ can then be described to a good approximation as a
c-number, $\hat{c}(0)\to \sqrt{N}\exp(i|\Omega_0|^2t/4\Delta)$, where
$N$ is the mean number of atoms in the condensate. This
approximation neglects both the depletion which occurs as atoms are
transferred into the side-modes $q \neq 0$ and the cross-phase modulation
between the condensate and the probe field, thus it is valid for times
short enough that $\sum_{{\bf q}\neq 0}\langle\hat{c}^\dag({\bf q})
\hat{c}({\bf q})\rangle
\ll N$, and $\langle \hat{A}^\dag\hat{A}\rangle \ll |\Omega_0|^2/4|g|^2$.
This is the matter-wave optics analog of the familiar classical and undepleted
pump approximation of nonlinear optics. Hence we describe the optical and
matter-wave fields on equal footings, treating all strongly populated modes
classically and all weakly populated modes quantum-mechanically.

Once we have replaced the condensate mode with its c-number counterpart, we
then neglect all terms in the Hamiltonian (\ref{H}) involving the product of
three or more weakly populated field modes. This is
a direct consequence of Bose enhancement, which strongly strengthens the
interactions involving the central mode $q=0$ mode relative
to those involving only the side-modes, and leads us to the effective
Hamiltonian
\begin{eqnarray}
\hat{H}&=&\hbar\omega_r\left[\hat{c}^\dag_+\hat{c}_+
+\hat{c}^\dag_-\hat{c}_--\delta\hat{a}^\dag\hat{a}\right.\nonumber\\
&+&\left.\chi\left(\hat{a}^\dag\hat{c}^\dag_-
+\hat{a}^\dag\hat{c}_++\hat{c}^\dag_+\hat{a}+\hat{c}_-\hat{a}\right)\right],
\label{3mH}
\end{eqnarray}
where $\omega_r=\hbar K^2/2m$, and
we have introduced the slowly varying operators
$\hat{c}_\pm=\exp(i|\Omega_0|^2t/4\Delta)\hat{c}(\pm {\bf K})$
and $\hat{a}=-i(g^\ast\Omega_0^\ast\Delta/|g||\Omega_0||\Delta|)
\exp(i\omega_0t)\hat{A}$.
The system is fully characterized by the effective coupling constant
$\chi=|g||\Omega_0|\sqrt{N}/8\omega_r\Delta$ and the dimensionless pump-probe
detuning $\delta=(\omega_0-\omega)/\omega_r$.

The Hamiltonian (\ref{3mH}) describes three coupled field modes:
the optical probe and two atomic condensate side-modes with wavenumbers
$\pm {\bf K}$.
The presence of terms such as $\hat{a}^\dag\hat{c}^\dag_-$ in Eq. (\ref{3mH})
is a direct consequence of momentum conservation, and does {\em not}
result from keeping antirotating terms in the Hamiltonian.
It immediately brings to mind
the optical parametric amplifier \cite{Walls}, a device
known to generate highly non-classical optical fields exhibiting
two-mode intensity correlations and squeezing, and which has been extensively
employed in the creation of entangled photon pairs for fundamental studies of
quantum mechanics, quantum cryptography and quantum computing.
A novel aspect of the present system is that it offers a way
to achieve quantum entanglement not just between optical fields, but rather
between macroscopically populated atomic Schr\"odinger and optical fields.

The dynamics of the system can be determined by solving the three
coupled-mode equations
\begin{equation}
\frac{d}{d\tau}
\left(\begin{array}{c}
\hat{d}_a\\ \hat{d}_-\\ \hat{d}_+\\ \end{array}\right)
=i\left(
\begin{array}{ccc}
\delta&-\chi&-\chi\\
\chi&1&0\\
-\chi&0&-1\\
\end{array}\right)
\left(\begin{array}{c}
\hat{d}_a\\ \hat{d}_-\\ \hat{d}_+\\ \end{array}\right),
\label{eqs}
\end{equation}
where $\tau=\omega_rt$, and we have introduced $\hat{d}_a(\tau)\equiv
\hat{a}(\tau)$, $\hat{d}_-(\tau)\equiv\hat{c}^\dag_-(\tau)$,
and $\hat{d}_+(\tau)\equiv\hat{c}_+(\tau)$ for future notational compactness.

An analytic solution can be constructed explicitly from the eigenvalues
$\{\lambda_j\}$ and eigenvectors $\{{\bf v}_{j}\}$ of the matrix on the
right-hand side of (\ref{eqs}). The eigenvalues $\{\lambda_j\}$
have been studied in detail in Ref. \cite{MooMey98b} in the context
of the theory of the Collective Atom Recoil Laser (CARL). It was
shown that provided the system parameters $\chi$ and $\delta$
satisfy certain threshold conditions, they take the form
$\lambda_1=\omega_1$, $\lambda_2=\lambda^\ast_3=\Omega+ i\Gamma$, where
$\omega_1$, $\Omega$, and $\Gamma$ are all real quantities. Hence we see
that after an initial transient, the solution grows exponentially in time
at the rate $\Gamma$. This regime of exponential growth is familiar
from the physics of the free-electron laser and of the CARL, where it
is usually studied at high temperatures. The explicit form of the eigenvalues and
eigenvectors is not required for the current analysis and will be presented
elsewhere. For our present purposes, it is sufficient to know that
for a given set of parameters $\chi,\delta$ they are simply constants.

The solution of Eqs. (\ref{eqs}) is
\begin{equation}
{\hat d}_i(\tau) = \sum_j u_{ij}(\tau) {\hat d}_j(0) ,
\end{equation}
where the coefficients $u_{ij}(\tau)$ are given by
\begin{equation}
u_{ij}(\tau)=\sum_k v_{ik}v^{-1}_{kj}e^{i\lambda_k\tau}
\approx \zeta_{ij}e^{(\Gamma+i\Omega)\tau}.
\label{defuij}
\end{equation}
Here ${\bf V}^{-1}$ is the inverse of the matrix ${\bf V}$ whose elements
$v_{ij}$ are given by the $i$th component of the eigenvector ${\bf v}_j$,
and $\zeta_{ij} \equiv v_{i3}v^{-1}_{3j}$. The approximate equality in
Eq. (\ref{defuij}) is valid for times long enough that one
can neglect all but the exponentially growing terms, henceforth referred to
as the exponential growth regime.

This exponential growth of the system can be triggered either from vacuum
fluctuations, as we discuss in more detail shortly, or by a weak
injected probe signal. We investigate both situations by assuming that the
probe field is initially in the coherent state $\alpha$, the vacuum state
corresponding to $\alpha=0$. The
condensate side-modes, in contrast, are always taken to be in the vacuum state
at $\tau=0$, so that the initial state of the system is $|\alpha, 0,0\rangle$.

The expectation values $\langle {\hat d}_i \rangle$ of the three coupled
modes are readily found to be
\begin{equation}
\langle\hat{d}_i(\tau)\rangle=\alpha u_{ia}(\tau)\approx \alpha
\zeta_{ia}e^{(\Gamma+i\Omega)\tau},
\label{meanfields}
\end{equation}
where the approximate result is for the exponential growth regime. As
expected, in the absence of an injected signal the mean fields remain
zero, but the injected probe breaks the symmetry of the system and lead to
nonzero expectation values.

Decomposing the expectation values $\langle d_i \rangle$ in terms of an
amplitude and phase as $\langle d_i \rangle = \ell_i(\tau)\exp[i\phi_i(\tau)]$,
we find that in the exponential growth regime their uncertainties obey
\begin{equation}
\frac{\Delta \ell_i(\tau)}{\ell_i(\tau)}
=\Delta\phi_i(\tau)\approx \frac{f(\chi,\delta)}{\sqrt{2}|\alpha|}
\label{dphi}
\end{equation}
where the fluctuation function $f(\chi,\delta)=|v^{-1}_{3-}/v^{-1}_{3a}|$
has a relatively simple dependence on the control parameters $\chi$ and
$\delta$. Specifically, for a given $\chi$, $f(\chi,\delta)$ is approximately
unity at the $\delta$ which maximizes the growth rate $\Gamma$, and
increases steadily away from this value. Clearly, Eq. (\ref{dphi})
holds only in the case of an injected probe signal, $\alpha \neq 0$.
In that case, the phase uncertainties of all three mean fields approach the
same limiting value for large $\tau$, and this value approaches zero as
$\alpha$ becomes very large, i.e. for large enough $\alpha$ all three modes
are effectively in coherent states. We note that for large $\alpha$ the
system is essentially equivalent to Kapitza-Dirac atomic diffraction of a 
condensate by a standing wave \cite{NIST}, which we now see produces atomic
side-modes in coherent states.

We now investigate the mean intensities $I_i(\tau)\equiv
\langle d_i^\dag(\tau) d_i(\tau) \rangle-\delta_{i,-}$. The
$\delta$-function accounts for the fact that $\hat{d}_2(\tau)$
in Eq. (\ref{eqs}) is a creation rather than an annihilation operator,
thus guaranteeing that the initial intensity of the``-'' side-mode vanishes.
These intensities are given explicitly by
\begin{equation}
I_i(\tau)=|\alpha|^2|u_{ia}(\tau)|^2+|u_{i-}(\tau)|^2-\delta_{i,-}.
\label{Ii}
\end{equation}
In the exponential growth regime they reduce to
\begin{equation}
I_i(\tau)\approx(|\alpha|^2|\zeta_{ia}|^2+|\zeta_{i-}|^2)e^{2\Gamma\tau}.
\label{Iilimit}
\end{equation}
They have a stimulated component, proportional to $|\alpha|^2$,
and a spontaneous component, which is present even when all three field
modes begin in the vacuum state. The stimulated component is simply
the squared amplitude of the mean field, while the spontaneous component
has no mean field, as it originates from the amplification of vacuum 
fluctuations in the atomic bunching.

To help understand this in more detail, we introduce the atomic ``bunching 
operator'' $\hat{B}=(1/N)\sum_j\exp(iK\hat{z}_j)$,
where $\hat{z}_j$ is the position operator of the $j$th atom. If the atoms
in the sample are evenly distributed in space then $\langle {\hat B} \rangle
= 0$. At the opposite extreme, if all the atoms are localized on a
array of period $2\pi/K$, then $|\langle {\hat B} \rangle|=1$.
Second-quantizing $\hat{B}$ and linearizing the result by treating
the $q=0$ mode as a c-number and keeping as in the derivation of Eq. (\ref{3mH})
only the lowest order terms in the side-mode operators, we can
reexpress ${\hat B}$ in terms of the atomic field operators as
$\hat{B}=(1/\sqrt{N})(\hat{c}^\dag_-+\hat{c}_+)$.
It is immediately apparent from that definition that
$\langle {\hat B}(0)\rangle=0 $ for our intial state $|\alpha, 0,0 \rangle$.
However, the
fluctuations $\langle {\hat B}^2(0) \rangle$ are nonzero, due to the
fact that $\langle\hat{c}_-(0)\hat{c}^\dag_-(0)\rangle=1$.
It is precisely this expectation value which leads to the spontaneous intensity
component, which can therefore be attributed to vacuum fluctuations
in the initial atomic bunching. These fluctuations play a role similar to
that of vacuum fluctuations in spontaneous emission.

In addition to the side-mode intensity, it is instructive to also study
their equal-time intensity correlation functions, as they provide useful
information about their quantum or classical nature. For the probe mode,
we have
\begin{equation}
g^{(2)}_a(\tau)=\frac{
\langle\hat{a}^\dag(\tau)\hat{a}^\dag(\tau)\hat{a}(\tau)\hat{a}(\tau)\rangle}
{\langle\hat{a}^\dag(\tau)\hat{a}(\tau)\rangle^2},
\label{defg2}
\end{equation}
The side-mode correlation functions $g^{(2)}_-(\tau)$ and $g^{(2)}_+(\tau)$
are defined likewise but with $\hat{a}(\tau)$ replaced by
$\hat{c}_-(\tau)$ and $\hat{c}_+(\tau)$ respectively.
These correlation functions are given explicitly as
\begin{equation}
g^{(2)}_i(\tau)=2-\frac{{\alpha}^2|u_{ia}(\tau)|^4}{I_i^2(\tau)}
\approx 2-\frac{|\alpha|^4}{[|\alpha|^2+f(\chi,\delta)]^2}.
\label{g2}
\end{equation}
As before, the approximate result applies to the exponential growth regime,
where the intensity correlation functions become constant in time
and the same for each mode.
In the case where the system builds up from noise ($|\alpha|^2 = 0$), we
have then $g^{(2)}=2$, the signature of a thermal or chaotic field.
As the injected signal strength is increased, however, $g^{(2)} \to
1$, which is characteristic of a Glauber coherent field with Poissonian
excitation statistics. Note the important point that the state of the
side-modes can be continuously varied from thermal to coherent by varying
the strength of the injected probe signal and/or the system parameters
$\chi$ and $\delta$. This demonstrates that the state of coherence of
the matter-wave field can be directly controled by an optical field, a new
form of coherent control.

We have mentioned the analogy between the problem at hand
and the parametric oscillator. It is the tool of choice for generating
entangled quantum optical states.
We now investigate if similar entanglements can be obtained here. 
We proceed by investigating the equal-time two-mode intensity
cross-correlations, which are a measure of the degree of entanglement
between the modes of the system. For example, the intensity cross-correlation
function $g^{(2)}_{a-}(\tau)$ is defined as
\begin{equation}
g^{(2)}_{a-}=\frac{\langle\hat{a}^\dag(\tau)\hat{a}(\tau)
\hat{c}^\dag_-(\tau)\hat{c}_-(\tau)\rangle}
{\langle\hat{a}^\dag(\tau)\hat{a}(\tau)\rangle
\langle\hat{c}^\dag_-(\tau)\hat{c}_-(\tau)\rangle}.
\label{defg212}
\end{equation}
Other intensity cross-correlation functions such as $g^{[2]}_{a+}(\tau)$ and
$g^{(2)}_{-+}(\tau)$ are defined similarly.

For classical fields, there is an upper limit to the second-order
equal-time correlation function. It is given by the Cauchy-Schwartz inequality
\cite{Walls}
\begin{equation}
g^{(2)}_{ij}(\tau)\leq \left[g^{(2)}_i(\tau)\right]^{1/2}
\left[g^{(2)}_j(\tau)\right]^{1/2}.
\label{CS}
\end{equation}
Quantum mechanical fields, however, can violate this inequality
and are instead constrained by the inequality \cite{Walls}
\begin{equation}
g^{(2)}_{ij}(\tau)\leq\left[g^{(2)}_i(\tau)+\frac{1}{I_i(\tau)}\right]^{1/2}
\left[g^{(2)}_j(\tau)+\frac{1}{I_j(\tau)}\right]^{1/2},
\label{qmineq}
\end{equation}
which reduces to the classical result in the limit of large intensities.

We focus our attention on the spontaneous case $\alpha=0$, where 
the single-mode intensity correlation functions are given by
$g^{(2)}_i(\tau)=2$. In this case,
the equal-time intensity cross-correlation functions are found to be
\begin{eqnarray}
g^{(2)}_{a-}&=& g^{(2)}_{-+} =
\left[2+\frac{1}{I_a(\tau)+I_+(\tau)}\right]^{1/2}
 \left[2+\frac{1}{I_-(\tau)}\right]^{1/2}, \nonumber \\
g^{(2)}_{a+} &=& 2.
\label{g212}
\end{eqnarray}
From Eq. (\ref{g212}) we see that both $g^{(2)}_{a-}(\tau)$ and
$g^{(2)}_{-+}(\tau)$ violate the Cauchy-Schwartz inequality, while
$g^{(2)}_{a+}(\tau)$ is consistent with classical cross-correlations.
Furthermore, the explicit evaluation of the $\zeta_{ij}$'s shows that
$I_+(\tau)\ll I_a(\tau)$, which implies that $g^{(2)}_{a-}(\tau)$ is very
close to the maximum violation of the classical inequality consistent
with quantum mechanics, whereas for $g^{(2)}_{-+}(\tau)$
the violation is not as strong. In the two-mode parametric amplifier,
the two-mode correlation function shows the maximum violation of the
Cauchy-Schwartz inequality consistent with quantum mechanics. In the three-mode
system, however, the two-mode cross-correlation functions involve a trace
over the third mode, hence it is not surprising that the two-mode correlations
are not maximized.

If we now allow for an injected coherent
probe field $(\alpha\neq 0)$, we must first note that the intensities
are increased by approximately $|\alpha|^2$, which means that the time scale 
on which the classical and quantum upper limits (\ref{CS}) and
(\ref{qmineq}) converge is reduced by $1/|\alpha|^2$,
making an experimental confirmation of quantum correlations more difficult.
In addition, whereas
for the spontaneous case $\alpha = 0$, numerics show the cross-correlation
$g^{(2)}_{a-}$ follows almost exactly the quantum upper limit (\ref{qmineq})
for all $t>0$, for $\alpha \neq 0$, it lies somewhere in between the quantum
(\ref{qmineq}) and classical (\ref{CS}) limits. As $\alpha$ is increased,
it falls ever closer to the
classical upper limit, so that in the limit of very large $\alpha$,
the fields exhibit only classical cross-correlations.

In summary, we have discussed how the quantum state
of momentum side-modes of a condensate can be varied continuously
between two distinct limits by specifying the initial state of an
optical cavity mode. When it begins in the vacuum state,
the side-mode and the cavity mode fields develop with zero mean fields, thermal 
intensity fluctuations, and strong quantum correlations between the modes. 
In contrast, when it is prepared in strong coherent state, we approach a 
``classical'' limit in which 
the fields develop with non-zero mean fields having well defined phases, intensity 
fluctuations indicating a coherent state, and exhibiting classical 
correlations only. 
Condensate side-modes have recently been realized in an experiment at NIST,
where a condensate was subjected to Kapitza-Dirac diffraction by a
standing-wave laser field.\cite{NIST} In order to observe the effects predicted
here, this field would need to be replaced by a combination of a strong
pump laser and a weak counter-propagating probe sustained by an optical
cavity, which does not appear to present any major difficulty.

This work has been supported in part by the U.S. Office of Naval Research
Contract No. 14-91-J1205, by the U.S. Army Research Office and by the
Joint Services Optics Program. P. M. acknowledges partial support by the
Humboldt Foundation.

\end{document}